\documentstyle[]{mn0}

\def\siml{\mathrel{\rlap{\raise 0.511ex \hbox{$<$}}{\lower 0.511ex \hbox{$\sim$}}}}
\def\Eg{{\cal E}_\gamma} \def\Egp{{\cal E}'_\gamma} \def\Ek{{\cal E}_k} 
\def\amati{$E_p \propto {\cal E}_\gamma^{1/2}$~} 

\begin{document}

\title[\amati for external-shock emission]
      {An external-shock origin of the \amati relation for Gamma-Ray Bursts}

\author[A. Panaitescu]{A. Panaitescu \\
       Space Science and Applications, MS D466, Los Alamos National Laboratory,
       Los Alamos, NM 87545, USA}

\maketitle

\begin{abstract}
\begin{small}
 We investigate the possibility that the \amati relation between the peak energy 
$E_p$ of the $\nu F_\nu$ spectrum and energy output $\Eg$ for long-duration GRBs 
arises from the external shock produced by the interaction of a relativistic 
outflow with the ambient medium. To that aim, we take into account the dependence 
of all parameters which determine $E_p$ and $\Eg$ on the radial distribution of
the ambient medium density and find that the \amati relation can be explained 
if the medium around GRBs has a universal radial stratification.
For various combinations of GRB radiative process (synchrotron or inverse-Compton) 
and dissipation mechanism (reverse or forward shock), we find that the circumburst 
medium must have a particle density with a radial distribution different than the 
$R^{-2}$ expected for constant mass-loss rate and terminal speed.
\end{small}
\end{abstract}


\section{Introduction}

 Lloyd, Petrosian \& Mallozzi (2000) have established that the 25--1000 MeV fluence 
$\Phi$ of bright BATSE Gamma-Ray Bursts (GRBs) is strongly correlated with the photon 
energy $E_p^{(obs)}$ at which peaks the burst $\nu F_\nu$ spectral energy distribution. 
More recently, Sakamoto et al (2008) have shown that 83 Swift-BAT and HETE-2 bursts 
display a $E_p^{(obs)} \propto \Phi^{0.52\pm 0.11}$ correlation, with the burst fluence 
measured at 15--150 keV, while Ghirlanda et al (2008) report $E_p^{(obs)} \propto 
\Phi^{0.32\pm 0.05}$ for 76 bursts (with known redshifts), the burst fluence being
calculated in the 1 keV--10 MeV range (i.e. bolometric).

 Lloyd et al (2000) found that the 8 GRBs with redshifts known at that time are not
standard candles and, thus, the $E_p^{(obs)}-\Phi$ correlation is not due to cosmological 
effects but is, most likely, intrinsic. In that venue, Amati et al (2000) have shown 
that the intrinsic peak energy $E_p$ and the isotropic-equivalent burst output $\Eg$ 
at $1-10^4$ keV are correlated, $E_p \propto \Eg^{0.52 \pm 0.06}$, for a set of 9 bursts 
with known redshifts (most of which are among those used by Lloyd et al 2000). 
Later, Amati (2006) found that $E_p \propto \Eg^{0.49 \pm 0.06}$ for a set of 41 GRBs,
while Ghirlanda et al (2008) arrive at $E_p \propto \Eg^{0.54 \pm 0.01}$ for 76 bursts. 
 
 The lack of bursts with a high fluence and average/low peak energy bursts in the $E_p^{(obs)} 
- \Phi$ correlation is, evidently, not due to selection effects (i.e. at least half 
of that correlation is real), with the thresholds for burst triggering and measuring
the peak energy possibly affecting only bursts with a high peak energy and low/average 
fluence. Ghirlanda et al (2008) and Nava et al (2008) investigate this possibility and
conclude that selection effects are negligible for pre-Swift bursts but do truncate
the distribution of Swift bursts in the $E_p^{(obs)}-\Phi$ plane. However, as the range
of peak energies of Swift bursts is much narrower than that of the entire sample, they
conclude that the $E_p^{(obs)}-\Phi$ is not an artifact of selection effects. 


\section{Possible origins for the \amati relation}

 The simplest explanation of the \amati relation is that it arises from viewing 
geometry and/or relativistic effects. Such an explanation is generic, i.e. it does
not make use of a certain mechanism for the production of the GRB. 

 In the former framework, GRBs arise from narrow jets seen at various angles $\theta$, 
the intrinsic burst emission being relativistically boosted by a factor 
$D = [\Gamma (1-\beta\cos\theta)]^{-1} \simeq 2/(\Gamma \theta^2)$, with $\Gamma$ 
being the jet Lorentz factor and the viewing angle $\theta$ being larger than the 
both the jet opening and the relativistic beaming angle $\Gamma^{-1}$. Relativistic 
beaming of the comoving frame emission (denoted with primed quantities) implies 
that the observed burst peak energy is $E_p = D E'_p$ and the inferred 
isotropic-equivalent GRB output is $\Eg = D^3 \Egp$ (the factor $D^3$ arising 
from $D^2$ for angular beaming and $D$ for boost of photon energy). 
Hence, in this scenario, the simplest expectation is that $E_p \propto \Eg^{1/3}$,
assuming that the comoving-frame peak energy $E'_p$ and GRB output $\Egp$ are 
{\sl universal} (i.e. they have same values for all bursts) or, at least, uncorrelated. 
Toma, Yamazaki \& Nakamura (2005) obtain the $E_p \propto \Eg^{1/3}$ analytical 
expectation in a more sophisticated way (for a typical GRB spectrum) but their 
numerical integration of the Doppler-boosted emission yields $E_p \propto \Eg^{0.4}$ 
for observer offsets that are comparable to (but larger than) the jet opening, 
which corresponds to higher energies $E_p$ and $\Eg$. 
Using an annulus geometry for the GRB outflow (i.e. a hollow jet), Eichler \& Levinson 
(2004) obtain a relation between the apparent $E_p$ and $\Eg$ consistent with or 
shallower than the \amati relation.

 A potential problem with the off-aperture jet model is the expected distribution 
of GRB peak photon fluxes. The average photon flux (taken as a measure for the peak
photon flux) of bursts seen at an offset angle less than $\theta$ is larger than 
$C(\theta) \propto \Eg / (E_p t_\gamma) \propto D^3 \Egp /(DE'_p t'_\gamma/D) 
\propto D^3 \propto \theta^{-6}$, where $t_\gamma$ is the burst duration. The number 
of such bursts is $N(<\theta) \propto \theta^2$. Thus, the cumulative peak-flux 
distribution expected in this model is $N(>C) \propto C^{-1/3}$ (for a volume-limited
sample), which is flatter than that measured by BATSE (e.g. Pendleton et al 1996), 
showing $N (>C) \propto C^{-1}$ at peak fluxes between 1 and 10 photons/${\rm cm^2\,s}$ 
and $N (>C) \propto C^{-3/2}$ at peak fluxes above 10 photons/${\rm cm^2\,s}$.

 Relativistic beaming of the emission from a jet wider than $\Gamma^{-1}$ and seen from
a location within its aperture may also be a possible origin of the \amati relation, 
as both quantities of interest, $E_p$ and $\Eg$, are affected by the source relativistic 
motion. In this case, $D \simeq \Gamma$, $E_p = \Gamma E'_p$, and $\Eg = \Gamma \Egp$
(only one power of $D$ because relativistic angular beaming also reduces the source
observed angular size by a factor $D^2$ relative to that of the entire source), hence 
$E_p \propto \Eg$ is expected if comoving-frame burst properties were universal or
uncorrelated. Thus, the \amati relation cannot be explained with just relativistic 
effects and requires a correlation of the comoving-frame peak energy $E'_p$ and GRB 
output $\Egp$ or a correlation of at least one of these quantities with the source
Lorentz factor $\Gamma$ (Schaefer 2003). Evidently, progress in this direction requires 
that a specific mechanism for the GRB emission generation is adopted (as done below).

 In that venue, Zhang \& M\'esz\'aros (2002) showed that the \amati relation may be 
accommodated with internal shocks, by noting that the comoving-frame magnetic field 
of a Poynting outflow (or that generated through shock dissipation) satisfies 
$B \propto L_{p/int}^{1/2}/R\Gamma$, where $L_{p/int}$ is the outflow's Poynting 
flux luminosity (or that of the dissipated, internal energy) and $R$ is the radius 
at which the burst emission is produced. The GRB synchrotron emission peaks at 
$E_p \propto \gamma^2 B \Gamma \propto \gamma^2 L_{p/int}^{1/2}/R$, where $\gamma 
m_e c^2$ is the typical electron energy in the GRB source. Thus, one obtains the \amati
relations if (1) the outflow's Poynting (or internal energy) luminosity is a good 
measure of the GRB output (in the sense that the $L_{p/int}/\Eg$ ratio is universal) 
and if (2) $\gamma$ and $R$ are universal (or not correlated with $L_{p/int}$). 

 Note that the above argument applies to any dissipation mechanism. For internal shocks,
the first condition above would lead to a constraint between the history of ejecta
Lorentz factors ($\Gamma(t)$) and the distribution of ejecta mass with the Lorentz 
factor, while the second requirement for the GRB radius would constrain only $\Gamma(t)$. 
As for the condition on the electron Lorentz factor, we note that, if electrons acquire 
a fraction $\epsilon_e$ of the outflow's internal energy, i.e. $N \gamma m_e c^2 = 
\epsilon_e U'$, where $N \propto \Ek/\Gamma$ is the electron number ($\Ek$ being the outflow
isotropic-equivalent kinetic energy) and $U' \propto V'u' \propto V'L (R\Gamma)^{-2}$ 
is the internal energy (with $V' = 4\pi R^2 ct'_\gamma$ being the volume of the GRB source), 
then $\gamma \propto \epsilon_e L_{p/int}/L_k$, where $L_k = \Ek/t_\gamma$ is the outflow 
kinetic luminosity. The requirement that $\gamma$ is not correlated with $L_{p/int}$ 
(leading to the \amati relation) implies that either $\epsilon_e \propto L_{p/int}^{-1}$ 
or that $L_k \propto L_{p/int}$, otherwise one would obtain that $E_p \propto L_{p/int}^{5/2}
\propto \Eg^{5/2}$.  

 Similarly, constraints on some model properties are required to explain the \amati
relation if the burst emission results from Comptonization of the thermal radiation
produced by magnetic reconnection or shock dissipation below the baryonic and/or 
pair photospheres. In this model (M\'esz\'aros \& Rees 2000, Ryde 2004, Ramirez-Ruiz 2005), 
the GRB peak energy and luminosity are correlated because both depend on the photospheric 
temperature. Rees \& M\'esz\'aros (2005) have shown that, if dissipation occurs above 
the saturation radius, then $E_p \propto \Gamma^2 L_\gamma^{-1/4}$, where $L_\gamma$ is 
the GRB luminosity. Then the \amati relation requires a certain correlation of the burst 
luminosity with the photosphere's Lorentz factor. Within the same model for the burst
emission, Thompson (2006) has shown that the \amati relation is obtained if the burst 
thermal radiation is produced at the stellar progenitor's photosphere, for which the
rest-frame temperature of the photons is $T'_{bb} \propto (L_\gamma/\Gamma^2)^{1/4}$,
and assuming that the outflow opening is set by its lateral expansion at the sound speed
($\theta \propto \Gamma^{-1}$) and that the collimation-corrected GRB output ($\propto
L_\gamma \theta^2$) is universal (as was first indicated by the analysis of Frail et
al 2001 and later shown to be incorrect by that the GRB collimated output ranges 
over 2 decades -- e.g. figure 1 of Ghirlanda, Ghisellini \& Lazzati 2004).  

 In this work, we present a possible origin of the \amati relation related to the dynamics 
of the GRB source, assuming an observer located within the opening of the relativistic 
outflow (i.e. viewing geometry is not at work). If the burst emission is synchrotron, 
then the peak of the GRB $\nu F_\nu$ spectrum is at $E_p \propto \gamma^2 B \Gamma$ and 
the flux density at that photon energy is $F_p \propto B \Gamma N$, where $\gamma$ is the 
electron typical comoving-frame Lorentz factor and $N$ the number of electrons in the 
GRB source. The GRB output being $\Eg \sim F_p E_p t_\gamma$, it follows that the \amati 
relation requires
\begin{equation}
 B \Gamma \gamma^2 \propto B \Gamma \gamma (N t_\gamma)^{1/2} \;.
\label{sy}
\end{equation}
If the burst emission were inverse-Compton scatterings of the synchrotron emission
generated by same electrons, then $E_p$ picks an extra-factor $\gamma^2$ and $F_p$
a factor $\tau$, the optical thickness to electron scattering of the GRB source.
Then the \amati relation requires
\begin{equation}
 B \Gamma \gamma^4 \propto B \Gamma \gamma^2 (N t_\gamma \tau)^{1/2} \;.
\label{ic}
\end{equation}

 It is tempting to attribute the \amati relation to $(i)$ variations from burst to burst
of the $B\Gamma$ factor, which appears both in the peak energy $E_p$ and the GRB 
output $\Eg$, $(ii)$ universality of $\gamma$, and $(iii)$ the remaining "dummy" parameters 
($N$, $t_\gamma$, $\tau$) being either universal or uncorrelated with $B\Gamma$
(so that they do not yield a different $E_p-\Eg$ dependence). We note that variations 
in $\gamma$ (for synchrotron) or $\gamma^2$ (for inverse-Compton) from burst to burst
that are larger than those of $B\Gamma$ would induce a $E_p \propto \Eg$ correlation
for either emission process.

 The burst duration $t_\gamma$, which is the only observable that appears in equations 
(\ref{sy}) and (\ref{ic}), has a spread of 1.5-2.0 orders of magnitude among long-bursts, 
which is comparable to the observed spread in GRB energy $\Eg$ at fixed peak energy $E_p$ 
(see figure 1 of Ghirlanda et al 2008). This suggests that the observed 
spread in the \amati relation requires some correlation among the dummy parameters, 
although it is possible that the range of measured $\Eg$ is smaller than the true spread 
because, for a fixed peak energy, bursts of a lower GRB output may fall below detection.

 If the GRB emitting electrons are accelerated at shocks, then it is unlikely that the 
product $B\Gamma$ can vary among bursts while $\gamma$ is universal, because acceleration
of electrons at relativistic shocks is expected to yield an electron Lorentz factor 
$\gamma$ that depends on that of the shock. As the latter bears a connection with 
the GRB source Lorentz factor $\Gamma$, a universal $\gamma$ requires either universal 
$\Gamma$ and $\epsilon_e$ or an ad-hoc correlation of these parameters. In the former case,
the \amati relation would rest entirely on variations in the magnetic field $B$ among 
bursts. The nearly 3 decades spread in observed $E_p$ and that $E_p \propto B \propto 
\epsilon_B^{1/2}$ imply that the fraction $\epsilon_B$ of the internal energy 
stored in shock-generated magnetic fields has a range of 6 decades.
Thus, a universal $\gamma$ requires a mechanism for electron acceleration at shocks 
that is completely decoupled from to the generation of magnetic fields, which is an
extreme requirement. For example, in the Weibel instability model of Medvedev (2006), 
proton current filaments created by the instability produce electric fields which 
accelerate electrons over distances of about the proton plasma skin-depth, leading to 
$\epsilon_e \simeq \epsilon_B^{1/2}$, hence the 3 decades range of observed peak energies
$E_p$ would be associated with an electron $\gamma$ which is far from being universal. 

 Thus, it seems much more likely that \amati relation is not due just to variations
in the $B\Gamma$ term among bursts and that some or all of the other parameters 
appearing in equations (\ref{sy}) and (\ref{ic}) contribute as well. To include their
effect in driving the \amati relation, we assume that the outflow's energy is dissipated 
by shocks which accelerates electrons and generate magnetic fields that acquire 
quasi-universal fractions of the dissipated energy. Some justification for the latter 
assumption is that, if the electron and magnetic parameters $\epsilon_e$ and $\epsilon_B$ 
were correlated as for the Weibel instability model, then their variations among bursts 
would induce a $E_p \propto \Eg^{3/4}$ correlation for synchrotron emission and $E_p 
\propto \Eg^{5/6}$ correlation for inverse-Compton.

  In the following section, we study the implications of equations (\ref{sy}) and
(\ref{ic}), representing the \amati relation, in the framework of {\sl external shocks}.
We note that this model has the drawback that the efficiency of the GRB emission should 
be small (below 10 percent) for those bursts with a large number (hundreds) of pulses 
(Sari \& Piran 1997). 
The same can be done for internal shocks which, as discussed above, will lead to 
constraints on the distribution of the ejecta Lorentz factor with mass and ejection
time (or variability timescale). A low GRB efficiency is also expected for this model 
(e.g. Kumar 1999).

\section{External-shock emission}

 The external shock driven by the interaction of the relativistic ejecta with the
burst ambient medium offers two possible GRB sources: the reverse shock, which 
energizes the ejecta, and the forward shock, which sweeps-up the circumburst medium.
Denoting by $\Gamma'$ the Lorentz factor of either shock as measured in the frame
of the incoming gas (the ejecta or the ambient medium), the shock jump conditions
lead to an internal energy density in the shocked gas that is $u' = (\Gamma'-1) n' 
m_p c^2$, where $n'= (4\Gamma'+3) n'_0$ is the comoving-frame particle density in 
the shocked fluid and $n'_0$ that in the unshocked gas. Thus, for a relativistic 
shock $\Gamma' \gg 1$), the typical electron Lorentz factor is $\gamma \propto \Gamma'$ 
and the magnetic field is $B \propto \Gamma n^{1/2}$, where $n$ is the ambient 
medium density and $\Gamma$ the laboratory-frame Lorentz factor of the shocked gas 
(i.e. the GRB source), the latter being valid also for the reverse shock because 
the contact discontinuity between the two shocked media is in hydrostatic equilibrium,
(i.e. pressure and internal energy density is the same behind both shocks).

 For a source moving at Lorentz factor $\Gamma$, the burst duration is $t_\gamma =
R_\gamma/\Gamma^2$, where $R_\gamma$ is the GRB source radius, which results from 
either the spread in the photon arrival time across the visible area of angular 
opening $\Gamma^{-1}$ or from the observer duration of the source travel-time up 
to radius $R$ (provided that the source is decelerating or accelerating slower
than $\Gamma \propto R^{1/2}$). Adding that the optical thickness to electron 
scattering is $\tau \propto N/R_\gamma^2$, the \amati relation of equations 
(\ref{sy}) and (\ref{ic}) becomes
\begin{equation}
 \Gamma'^2 \Gamma^2 n^{1/2} \propto \Gamma' \Gamma (n N R_\gamma)^{1/2} 
\label{syn}
\end{equation}
for synchrotron emission and
\begin{equation}
 \Gamma'^4 \Gamma^2 n^{1/2} \propto \Gamma'^3 \Gamma  N (n/R_\gamma)^{1/2} 
\label{inv}
\end{equation}
for inverse-Compton.

 Below, we investigate the conditions required for the synchrotron and inverse-Compton 
emissions from the reverse and forward shocks to accommodate the \amati relation
with the following simplifications: \\
(1) the burst emission is produced before the reverse shock crosses the ejecta shell,
i.e. before the deceleration of the external shock starts.
One reason for this restriction is that the shock-crossing radius offers a "milestone" 
in the evolution of the external shock that could be the GRB radius $R_\gamma$, while 
no such reference point exist after deceleration sets in. A second reason is that it 
would be unnatural for a decelerating external shock to radiate episodically, once during
the burst, until 100 s, and then again staring after 1000 s, during the afterglow,
as observed in the X-ray emission of a majority of Swift GRBs (O'Brien et al 2006, 
Willingale et al 2007),  \\
(2) the entire emitting fluid moves at the same Lorentz factor $\Gamma(R_\gamma)$ 
and is filled with the same magnetic field $B(R_\gamma)$, with the values taken at
the radius were the burst emission is released. 
As $E_p$, $F_pE_p$, and $\Gamma$ are power-laws in the shock radius, for a radially
extended burst emission, their burst-averaged ($\overline{E_p} = \int E_p dF_p/ \int dF_p$) 
and burst-integrated ($\int F_p E_p dt \propto \Eg$ with $dt=dR/\Gamma^2$) values 
have the same dependence on $R_\gamma$ as their values at $R_\gamma$, \\
(3) the distribution with energy of the shock-accelerated electrons is softer than 
$dN/d\gamma_e \propto \gamma_e^{-3}$ above the typical $\gamma$, so that the peak 
of $\nu F_\nu$ is for the typical $\gamma$ electrons and not at a higher random
Lorentz factor determined by electron cooling and/or acceleration, \\
(4) the electrons with the typical Lorentz factor $\gamma$ do not cool significantly 
during the GRB emission. We note that only a small of the BATSE bursts (Preece et al 2000) 
have the $F_\nu \propto \nu^{-1/2}$ spectrum below the peak energy $E_p$ expected in 
the opposite case. If the $\gamma$-electrons cool during the burst, the $\nu F_\nu$ 
spectrum still peaks at the synchrotron or inverse-Compton energy corresponding to 
$\gamma$ (i.e. the peak energy $E_p$ remains unchanged), but the flux at $E_p$ picks 
a multiplying factor $\gamma_c/\gamma \propto (\Gamma' \Gamma n R_\gamma)^{-1}$ for 
synchrotron and a factor $(\gamma_c/\gamma)^2$ for inverse-Compton, owing to that 
most electrons are at the cooling Lorentz factor $\gamma_c \propto \Gamma/(B^2 R_\gamma)$ 
for which the radiative cooling timescale is equal to the burst duration. 

 Thus, the treatment provided below is not sufficiently comprehensive and serves only
as an illustration of the conditions required for the external shock to account for 
the \amati relation.

\subsection{Dense ejecta (semi-relativistic reverse shock)}
 
 For the evolution of the ejecta--ambient medium interaction at times before
the reverse shock crosses the ejecta shell (i.e. before the standard deceleration
sets-in), the shock jump conditions can be used to derive a fourth-degree equation 
for the Lorentz factor $\Gamma$ of the shocked fluid (which is the same for both the 
shocked ejecta and the swept-up ambient medium). As shown by Panaitescu \& Kumar (2004), 
the solution of that equation is  
\begin{equation}
\Gamma \simeq \Gamma_0 \left[ 1 + 2 \Gamma_0 \left( n/n'_{ej} \right)^{1/2} \right]^{-1/2} \;
\label{G}
\end{equation}
where $\Gamma_0$ is the Lorentz factor of the unshocked ejecta and $n'_{ej}$ their 
comoving-frame density.

 In the $n'_{ej} \gg 4 \Gamma_0^2 n$ limit (thin and dense ejecta shell), 
equation (\ref{G}) leads to $\Gamma \siml \Gamma_0$, independent of the $n/n'_{ej}$ 
ratio, and to a mildly relativistic reverse shock of constant Lorentz factor $\Gamma'$. 
If $\Gamma$ and $\Gamma'$ do not change with radius, then equations (\ref{syn}) and 
(\ref{inv}) imply that the \amati relation is induced by a certain correlation of 
the ejecta Lorentz factor $\Gamma_0$ with the radius $R_\gamma$ where the GRB emission 
is released. 
We focus on the forward shock emission because the mildly relativistic reverse shock 
is unlikely to yield an emission spectrum peaking in the hard X-rays. For the 
forward shock, $\Gamma' = \Gamma \simeq \Gamma_0$ and $N \propto n R_\gamma^3 = 
R_\gamma^{3-s}$ for an ambient medium density stratified as $n \propto R_\gamma^{-s}$ 
with $s<3$, and $N \simeq const$ if $s>3$. 

 For synchrotron emission and $s < 3$, equation (\ref{syn}) requires $\Gamma_0^2 
\propto R_\gamma^{2-s/2}$. If $\Gamma_0$ were universal, this leads to an inconsistent 
solution $s=4$. Thus $\Gamma_0$ should vary among bursts, in which case the requirement 
imposed by the \amati relation is that the {\sl GRB emission is released at a radius that 
is correlated with the ejecta Lorentz factor}. 
 Further investigation can be done if $R_\gamma$ is determined in some way. 
The termination shock of the wind expelled by the GRB progenitor is the only milestone
expected in the evolution of the forward shock, though it is not evident how it could set 
the GRB radius; even that were achieved, the location of the termination shock should 
not be related to the ejecta initial Lorentz factor. Instead, we speculate that the
location where the forward-shock GRB emission is released is tied to the radius $R_+ 
\propto (\Ek/\Gamma_0^2)^{1/(3-s)}$ at which the reverse shock crosses the ejecta 
($\Ek$ being the ejecta kinetic energy), and after which the blast-wave is decelerated. 
Then and the \amati relation can be obtained if $s=10/3$, which is inconsistent with 
the starting assumption $s<3$. 

 For synchrotron emission and $s > 3$, equation (\ref{syn}) requires $\Gamma_0^2 
\propto R_\gamma^{1/2}$. Relating $R_\gamma$ to the shock-crossing radius $R_+$, 
a self-consistent solution ($s=3.5$) is found, for which $E_p \propto \Gamma_0^4 
R_\gamma^{-7/4} \propto \Gamma_0^{-3} \Ek^{1/4}$. 
 
 For inverse-Compton emission and $s < 3$, equation (\ref{inv}) requires $\Gamma_0^2 
\propto R_\gamma^{2.5-s}$. If $\Gamma_0$ were universal, the \amati relation would 
be accounted for by an ambient medium with $s=2.5$. In this case, $R_\gamma = R_+$
leads to $R_\gamma \propto \Ek^2$ and $E_p \propto R_\gamma^{-11/8} \propto \Ek^{-5/2}$. 
If $\Gamma_0$ is not universal, the \amati relation is obtained for $s=11/4$, 
leading to $E_p \propto \Gamma_0^6 R_\gamma^{-11/8} \propto \Gamma_0^{17} \Ek^{-11/2}$.
 For $s>3$, the \amati relation requires $\Gamma_0^2 \propto R_\gamma^{-1/2}$,
which for $R_\gamma = R_+$ leads to $s=2.5$, i.e. an inconsistent solution.

 Therefore, for a thin ejecta shell, the \amati relation can be explained with 
synchrotron emission from the forward shock if GRBs are produced at the radius 
where the reverse shock crosses the ejecta shell and if the ambient medium
around bursts has a universal $n \propto R^{-3.5}$ radial structure, bursts with 
higher peak energy $E_p$ and GRB output $E_\gamma$ resulting for lower ejecta
Lorentz factors $\Gamma_0$ or larger ejecta kinetic energies $\Ek$. The \amati 
relation can also be obtained with inverse-Compton emission if $n \propto R^{-2.5}$
and $\Gamma_0$ are universal, bursts with higher $E_p$ and $E_\gamma$ resulting
from a lower $\Ek$, or if $n \propto R^{-2.75}$ for all bursts if $\Gamma_0$ is 
not universal, a higher $E_p$ and $E_\gamma$ being obtained for a higher 
$\Gamma_0$ or lower $\Ek$.

\subsection{Tenuous ejecta (relativistic reverse shock)}
 
 In the $n'_{ej} \ll 4 \Gamma_0^2 n$ limit (thick and tenuous ejecta shell),
equation (\ref{G}) leads to $\Gamma = (\Gamma_0/2)^{1/2} (n'_{ej}/n)^{1/4} \ll \Gamma_0$ 
and to a relativistic reverse shock with $\Gamma' \simeq \Gamma_0/(2\Gamma) \gg 1$. 
Considering that the radial width of the ejecta shell increases linearly with its 
radius, the comoving-frame ejecta density is $n'_{ej} \propto (\Ek/\Gamma_0)/ R^3$. 
Then, for a ambient medium with radial density profile $n \propto R^{-s}$, we obtain 
that the Lorentz factor of the shocked gas evolves as $\Gamma \propto R^{(s-3)/4}$. 
Therefore, if $s>3$, the shocked gas motion is accelerated by the ram pressure of 
the incoming ejecta, starting from a value well below $\Gamma_0$ (and remaining below 
it at all times). For $s<3$, the GRB source is decelerating (but this deceleration 
is substantially slower than that after the reverse shock has crossed the ejecta shell).

 In the following investigation, we drop the dependence of two quantities of 
interest, $E_p$ and $E_\gamma$, on the ejecta Lorentz factor $\Gamma_0$, i.e. 
we assume it to be universal, and determine the stratification index $s$ that 
accommodates the \amati relation. In this case, bursts have different peak
energies and GRB outputs because their emission is produced at different radii
$R_\gamma$. If $R_\gamma$ is identified with the shock having crossed the entire 
ejecta shell, then the GRB radius is set by the ejecta kinetic energy and the
duration of the ejecta release, which is about the same as the observer frame
burst duration: $R_+ \propto (\Ek t_\gamma)^{1/(4-s)}$ (Panaitescu \& Kumar 2004).

 If $\Gamma_0$ were not universal, the \amati relation could be explained if the 
GRB radius $R_\gamma$ and $\Gamma_0$ satisfy a certain relation. Then, relating 
$R_\gamma$ with the shock crossing radius will lead to a certain correlation among 
$\Gamma_0$, $\Ek$, and $t_\gamma$, an avenue which we will not explore any further.

 The continuous injection of relativistic electrons (in the downstream region) 
of a Lorentz factor $\gamma_e \propto \Gamma'$ which changes with the outflow radius 
will lead to an electron population at $R_\gamma$ that has a power-law distribution 
with energy, $dN/d\gamma_e \propto \gamma_e^{-q}$. The effective index $q$ can be
calculated by first determining the medium structure parameter $s(q)$ that accounts 
for the \amati relation, then the dynamics of the shocked fluid $\Gamma(R)$ (which 
sets $\gamma_e$) and the derivative $dN/dR$ of the electrons number, from where 
$dN/d\gamma_e$ can be obtained and the loop is closed to find the exponent $q$
\footnote{If, at each instant, a power-law distribution with index $p$ is injected 
 downstream, the cumulative electron distribution will have the index $q$ if $q<p$ and 
 the index $p$ if $q>p$. For simplicity, we assume that the former is always the case.} .
Because $\gamma$ evolves with $R$, one must check that the assumed location of the 
peak of $\nu F_\nu$ is consistent with the evolution of $\gamma(R)$ and the inferred
effective index $q$ of the electron distribution.

\vspace{-3mm}
\subsubsection{Forward shock}

 For the forward shock, $\Gamma' = \Gamma \propto R_\gamma^{(s-3)/4}$ and
$N \propto R_\gamma^{3-s}$ if $s<3$, while $N \simeq const$ if $s>3$.
 For $s>3$, most of electrons have been accelerated before the GRB radius $R_\gamma$ 
and we have to find a self-consistent solution considering that the peak energy 
$E_p$ is either at $\gamma (R_\gamma)$ (which we will denote as $\gamma_1$) or at 
some electron Lorentz factor $\gamma_0$ corresponding to when relativistic electrons 
were first produced. We will make the simplifying assumption that $\gamma_0$ is 
a universal quantity. 

 For a $dN/d\gamma_e \propto \gamma_e^{-q}$ electron distribution, with most 
electrons being at $\gamma_0$, the flux at photon energy $E_0 \propto \gamma_0^2 
B\Gamma$ is $F_0 \propto \Gamma N B$, while the flux at energy $E_1 \propto 
\gamma_1^2 B\Gamma$ is $F_1 \propto \Gamma NB (\gamma_0/\gamma_1)^{q-1}$. 
(1) If the peak of $\nu F_\nu$ is at $E_0$, then it can be shown that the \amati 
relation requires $s=5$, for which $\gamma_e \propto R^{1/2}$, $dN/dR \propto 
R^{-3}$, hence $dN/d\gamma_e \propto \gamma_e^{-7}$. Given that $d\gamma_e/dR > 0$,
$q=7$ implies that the peak of $\nu F_\nu$ is, indeed, at $E_0$, consistent with
the starting assumption. For this case, $E_p \propto R_\gamma^{-3/2}$. \\
(2) If the peak of $\nu F_\nu$ is at $E_1$, then the \amati relation leads to 
$s=3+4/(q+3)$, $\gamma_e \propto R^{1/(q+3)}$, $q=5$ which, together with 
$d\gamma_e/dR > 0$, implies that the peak of $\nu F_\nu$ is, in fact, at $E_0$, 
in contradiction with the starting assumption.
 
 For synchrotron emission and $s<3$, the \amati relation given in equation 
(\ref{syn}) is satisfied if $s=3.5$, hence this is not a self-consistent solution. 

 For inverse-Compton emission and $s<3$, the \amati relation of equation (\ref{inv})
requires $s=19/7$, leading to $\gamma_e \propto R^{-1/14}$, $dN/dR \propto R^{-5/7}$,
and $q=5$ which, together with $d\gamma_e/dR <0$, implies that the peak of $\nu F_\nu$
is, indeed, determined by the $\gamma(R_\gamma)$ electrons. For this case, $E_p 
\propto R_\gamma^ {-25/16}$. 

 For inverse-Compton emission and $s>3$, the flux at photon energy $E_0 \propto 
\gamma_0^4 B\Gamma$ is $F_0 \propto \tau \Gamma N B$, while the flux at energy 
$E_1 \propto \gamma_1^4 B\Gamma$ is $F_1 \propto \tau \Gamma NB (\gamma_0/\gamma_1)^
{2(q-1)}$. \\
(1) If the peak of $\nu F_\nu$ is at $E_0$, the \amati relation requires $s=1$,
incompatible with the assumed $s>3$.\\
(2) If the peak of $\nu F_\nu$ is at $E_1$, then the \amati relation leads to 
$s=3-2/(q+2)$, $\gamma_e \propto R^{-1/(2q+4)}$, $dN/dR \propto R^{-q/(q+2)}$, 
from where $q=5$ and $s=19/7$, inconsistent with the $s>3$ initial assumption.

 Therefore, the \amati relation can be accommodated with the synchrotron emission
from the pre-deceleration forward shock if all GRBs occur in a $n \propto R^{-5}$ 
medium, but at different radii, or by with the inverse-Compton forward shock emission
if the ambient medium has a universal $n \propto R^{-19/7}$ stratification. In either
case, bursts of higher $E_p$ and $\Eg$ are those occurring at smaller radii.

\vspace{-3mm}
\subsubsection{Reverse shock}

 For the reverse shock, $\Gamma' \propto \Gamma^{-1} \propto R_\gamma^{(3-s)/4}$
and the number of energized ejecta electrons evolves as $dN/dR \propto R^2
(\Gamma_0 n'_{ej})(\beta_0 - \beta)$, where $\beta_0$ and $\beta$ are the 
lab-frame velocities of the unshocked and shocked ejecta, respectively.
For $\Gamma_0 \gg \Gamma$, we have $\beta_0 - \beta \simeq (2\Gamma^2)^{-1}$.
Using $n'_{ej} \propto R^{-3}$ and $\Gamma \propto R_\gamma^{(s-3)/4}$, one 
arrives at $dN/dR \propto R^{(1-s)/2}$, from where $N \propto R_\gamma^{(3-s)/2}$
for $s<3$ and $N \simeq const$ for $s>3$.

 For synchrotron emission and $s<3$, the \amati relation leads to a contradicting
$s=5$. For $s>3$, assuming that the $\nu F_\nu$ spectrum peaks at $E_0 \propto 
\gamma_0^2 B\Gamma$, the \amati relation requires $s=5$, implying $\gamma_e \propto 
R^{-1/2}$, $dN/dR \propto R_\gamma^{-2}$, thus $q=-1$ which, together with 
$d\gamma_e/dR < 0$, implies that the peak of $\nu F_\nu$ is, indeed, at $E_0$. 
For this case, we obtain $E_p \propto R_\gamma^{-3/2}$.
For $s>3$, assuming that the $\nu F_\nu$ spectrum peaks at $E_1 \propto \gamma_1^2 
B\Gamma$, the \amati relation requires $s=3-4/(q+1)$, from where $\gamma_e \propto 
R_\gamma^{1/(1-q)}$, $dN/dR \propto R^{(3-q)/(q-1)}$, leading to $q=3$ and $s=1$, 
inconsistent with the starting choice $s>3$.
 
 For inverse-Compton emission and $s<3$, the \amati relation requires $s=1$, 
yielding $\gamma_e \propto R^{1/2}$, $dN/dR = const$, hence $q=-1$, thus 
$d\gamma_e/dR > 0$ implies that $\nu F_\nu$ peak energy is determined by the
$\gamma_1$ electrons. In this case, $E_p \propto R_\gamma^{-1/2}$. 
For $s>3$, assuming that the $\nu F_\nu$ spectrum peaks at $E_0 \propto \gamma_0^4 
B\Gamma$, the \amati relation requires $s=1$, which is incompatible with the 
working condition $s>3$. 
For $s>3$, assuming that the $\nu F_\nu$ spectrum peaks at $E_1 \propto \gamma_1^4
B\Gamma$, the \amati relation leads to $s=3+2/q$, implying $\gamma_e \propto 
R^{-1/(2q)}$ and $dN/dR \propto R^{-1-1/p}$, from where $q=-1$ and $s=1$, again 
incompatible with the starting condition.

 Thus, synchrotron emission from the reverse shock can account for the \amati
relation provided that bursts occur at various radii in a $n \propto R^{-5}$
medium, while inverse-Compton emission can explain the same relation if 
$n \propto R^{-1}$. For either radiation process, bursts of higher $E_p$ and 
$\Eg$ are those occurring at smaller radii.

\section{Conclusions}

 We have investigated the ability of the external shock (produced by the interaction
of relativistic ejecta with the ambient medium) to accommodate the \amati relation
between the burst peak energy and isotropic-equivalent energy release.
First, we noted that it seems unlikely that the \amati relation is due to variations 
of the quantity $B\Gamma$ among bursts, with the electron Lorentz factor $\gamma$ being 
universal and all other quantities ($N$, $t_\gamma$, $\tau$) in the right-hand sides 
of equations (\ref{sy}) and (\ref{ic}) not being correlated with $E_p$. That is so 
because electron acceleration at relativistic shocks is likely related to the 
generation of magnetic fields and with the strength of the shock and because the 
number of radiating electrons $N$ and the burst duration $t_\gamma$ could be related 
with the dynamics of the external shock.
 
 For that reason, we have identified the conditions that lead to the \amati relation 
by taking into account all the quantities that determine $E_p$ and $\Eg$. 
After making some simplifications (uniform magnetic field, single Lorentz factor 
in the shocked fluid, negligible electron cooling), and considering only the external 
shock emission before the reverse shock crosses the ejecta shell (as an interrupted
burst--afterglow emission from the same decelerating outflow seems too contrived), 
we have determined the dependence of $E_p$ and $\Eg$ on the radius where (or up to which)
the burst emission is produced, with allowance for both the reverse and forward shock, 
synchrotron and inverse-Compton emissions, and relativistic or semi-relativistic reverse
shock. In the latter case, only the forward shock is expected to produce the high-energy
prompt emission, the \amati relation requiring a correlation between the ejecta Lorentz
factor and the GRB radius which does not have a plausible justification. For that
reason, we favour an explanation of the \amati relation where the reverse shock is 
relativistic. 

 In our treatment of that case, the burst emission is assumed to arise over a small 
range of source radii or up to a certain radius, the \amati relation resulting from 
variations in that radius from burst to burst. The reverse shock crossing the ejecta 
or the external shock encountering the termination shock of the progenitor's freely 
expanding wind are the milestones in the dynamical evolution of the reverse and 
forward shock, respectively, that could set the location where the burst radius is 
produced. This implies that the variations from burst to burst in the radius at 
which the prompt emission is released is due to either (1) variations among bursts 
in the kinetic energy of the ejecta or in the duration of ejecta release (for a reverse 
shock origin of the GRB), or (2) to the history of the mass-loss of the GRB progenitor 
shortly before its core collapse (for GRBs produced by the forward shock). 

 Within the external-shock model for GRBs, the \amati relation can be accounted for
by just the power-law radial stratification of the burst ambient medium density. 
For the four possible combinations of dissipation shock and radiation process, we find  
the following density profiles: $n \propto R^{-1}$ (inverse-Compton from reverse shock),
$R^{-19/7}$ (inverse-Compton from forward shock), and $R^{-5}$ (synchrotron from either
shock). In general, a steep ambient profile is required to explain the slope of the 
\amati relation because of the weak dependence of the source Lorentz factor on the 
density of the ambient medium ($\Gamma \propto n^{-1/4}$ -- equation \ref{G}).

 None of the ambient medium stratification required by the \amati relation is the 
$n \propto R^{-2}$ profile expected for a massive stellar GRB progenitor expelling 
a constant speed wind a steady mass-loss rate. Considering that the burst emission 
occurs at $\siml 10^{16.5}$ cm and that the wind termination shock moves at $\sim 
10 \; {\rm km\; s^{-1}}$), this implies that, in the last $\siml 1000$ years before 
core collapse, the Wolf-Rayet progenitor of long-bursts had a varying mass-loss 
rate or wind terminal velocity. 
 However, we do rule out that, by relaxing the simplifying assumptions made here,
the ambient medium density profile required to explain the \amati relation with the
external-shock emission becomes consistent with $n \propto R^{-2}$.

 As the burst model employed here is that of the external shock before the reverse
shock crosses the ejecta (i.e. before deceleration begins), the ensuing afterglow 
emission could be attributed to the emission from the reverse or forward  shocks 
after deceleration, with allowance for injection of ejecta and energy after the burst,
to account for the extended afterglow emission (if it is from the reverse shock) and 
the X-ray light-curve plateaus (if it is from the forward shock). Then, the general lack 
of continuity of burst-to-afterglow emissions, shown by the steep fall-off of the X-ray
flux by 2--3 dex at the end of the burst, would lead to a rather contrived model, 
where the discontinuous burst-to-afterglow emission requires a temporary switch-off 
of the external-shock emission, followed by a much softer emission (the afterglows).
A simpler is that where the two emission phases, prompt and delayed, are attributed 
to different outflows, with the burst arising from a narrower jet whose bright, 
high-energy emission is produced only before the reverse shock crosses the ejecta 
or the external shock reaches the wind termination shock, but having a sufficiently 
low, collimated kinetic energy, so that its post-burst (forward shock) emission is 
dimmer than that from a wider, more energetic outflow producing the afterglow emission.

\section*{Acknowledgments}
 The author acknowledges the support of the US Department of Energy through the 
LANL/LDRD 20080039DR program.

\end{document}